\documentclass[12pt]{article}
\usepackage{amsmath,amssymb,amsfonts,amsthm}
\newcommand{\be}{\begin{equation}}
\newcommand{\ee}{\end{equation}}
\newcommand{\bea}{\begin{eqnarray}}
\newcommand{\eea}{\end{eqnarray}}
\newcommand{\nn}{\nonumber \\}
\newcommand{\p}[1]{(\ref{#1})}
\newcommand{\lb}{\label}

\begin{document}

\begin{titlepage}

\vspace*{1.5cm}

\begin{center}
\begin{Large}
{\Large\bf Worldsheet Supersymmetry of Pohlmeyer-Reduced}
\vspace{0.5cm}

{\Large\bf $AdS_n\times S^n$ Superstrings}   \vspace{0.6cm}
\end{Large}

\vskip 0.3cm {\large {\sl }} \vskip 10.mm {\bf M.
Goykhman$^{\,1,\,2,\,a}$, $\;\;\;$  E. Ivanov$^{\,1,\,b}$

}
\vspace{1cm}

{\it
$^1$ Bogoliubov Laboratory of Theoretical Physics, JINR, \\
141980 Dubna, Moscow Region, Russia\\
$^2$ Department of General and Applied Physics, MIPT,\\
141700 Dolgoprudny, Moscow Region, Russia
}
\end{center}
\vfill

\par
\begin{center}
{\bf ABSTRACT}
\end{center}
\begin{quote}
As was observed by Grigoriev and Tseytlin, the Pohlmeyer-reduced
$AdS_2\times S^2$ superstring theory possesses ${\cal N}=(2,2)$
worldsheet supersymmetry. We show, at the classical level,  that the
$AdS_3\times S^3$ and $AdS_5\times S^5$ superstring theories  in the
Pohlmeyer-reduced form reveal hidden ${\cal N}=(4,4)$ and ${\cal
N}=(8,8)$ worldsheet supersymmetries. Our
consideration is based on the modified mass-deformed gauged WZW
action for the superstring equations. We present the explicit form
of the supersymmetry transformations for both the off-shell action
and the superstring equations. The characteristic feature of these
transformations is the presence of non-local terms.

\vfill \vfill \vfill \vfill \vfill \hrule width 5.cm \vskip 2.mm
{\small
\noindent $^a$ goykhman89@gmail.com\\
\noindent $^b$ eivanov@theor.jinr.ru\\
}
\end{quote}
\end{titlepage}

\section{Introduction}
In \cite{TG,GT} Grigoriev and Tseytlin (see also \cite{MS, RT, HIT, Iwa, HT})
applied the Pohlmeyer reduction method \cite{PO} for eliminating
non-dynamical degrees of freedom in GS type-IIB superstring on
$AdS_5\times S^5$ and GS superstring on $AdS_3\times S^3$ and
$AdS_2\times S^2\,$. In the latter case they demonstrated that the
eventual action possesses ${\cal N}=(2,2)$ worldsheet supersymmetry
and is none other than the action of ${\cal N}=(2,2)$ superextension
of combined sine-sinh-Gordon model. They also posed the question
about the appropriate worldsheet supersymmetries in the actions of
the Pohlmeyer-reduced (PR) $AdS_3\times S^3$ and $AdS_5\times S^5$
superstrings. To our knowledge, this question remained unanswered so
far.

In this paper we suggest a possible solution to the problem of worldsheet
supersymmetry of the PR $AdS_n\times S^n$ superstring action for the $n=3$ and $n=5$ cases.
Our proposal is based on several simple ideas.

Following \cite{TG, GT}, we adopt the supermatrix notation for the fields entering the action,
i.e. write the PR superstring action in the form maximally closed to that of supersymmetric
gauged WZW (gWZW) models (see \cite{Rohm, Witten-Matr, N-SgWSW} and refs. therein). New points as compared
to the formulation in \cite{TG, GT} are as follows.

First, we systematically use the Polyakov-Wiegmann \cite{Pol-Wig-1, Pol-Wig-2} type
representation for the gauge fields in
the generalized gWZW action, namely
\be
A_+ = u\partial_+ u^{-1}\,, \quad A_- = \bar u\partial_- {\bar u}^{-1}\,,\lb{1}
\ee
where $u$ and $\bar u$ are two independent matrices valued in the gauge group $H$. Due to this representation,
we obtain modified equations of motion for gauge fields\footnote{This substitution for $A_\pm$
was already used in \cite{TG} and \cite{RT}. However, our motivation is different.}.

Second, we modify the original gWZW action of \cite{TG, GT}
by adding the term which involves only the matrices  $u$ and $\bar u$ and is gauge invariant by itself:
\be
S_a = S^{(H)}_{WZW} (B)\,, \quad B = u^{-1}\bar u\,. \lb{2}
\ee
This addition does not influence the equations of motion for the physical matrix fields
$g, \Psi_{L, R}$ resulting from the Pohlmeyer reduction, but further modifies the equations
for the gauge fields.

With the special coefficient before the new term in the action, the
equations of motion for the gauge fields are automatically satisfied
as a consequence of those for physical fields and so do not impose
any restriction on the gauge fields at all. In fact, it is just the
value at which $S_{gWZW} + S_a = [S_{WZW}(u^{-1}g \bar u) -
S^{(H)}_{WZW}(B)] + S^{(H)}_{WZW}(B) = S_{WZW}(u^{-1}g \bar u)\,$.
Surprisingly, the same value of the coefficient before \p{2} is
required for off-shell supersymmetry\footnote{Hereafter, by
``off-shell'' we understand the supersymmetry of the action as
opposed to the possible supersymmetry of equations of motion. This
should not be confused with the more accustomed usage of this term
as meaning the realization of supersymmetry transformations on the
physical fields only, with the possible auxiliary fields being
eliminated by their equations of motion.}. The supersymmetry is
realized by the transformations which look similar to the
transformations, suggested in \cite{TG} as a generalization of those
for the $n=2$ model; however, they involve unremovable
non-localities and are free from some extra (too strong)
restrictions on the group parameters assumed in \cite{TG}.

We find the $(4,4)$-parameter chiral supersymmetries of the modified
action for the $n=3$ case and the $(8, 8)$-parameter chiral
supersymmetries for the $n=5$ case. We then derive an on-shell
closure of supersymmetries on the $2d$ worldsheet translations,
modulo some compensating gauge transformations. It still remains to
learn what the full off-shell superalgebras spanned by these odd
transformations (together with their bosonic closure) are.

In our notations we closely follow refs.~\cite{TG} and \cite{GT}; actually,
we take as an input the basic results of these papers, although some key
steps of the derivation of the PR superstring action are presented for completeness too.
We do this in Sect.~2. In Sect.~3 we pass to the modified gWZW action
with fermionic and potential terms giving rise to the same
PR superstring equations of motion as in \cite{TG, GT}.
Then we show that it possesses chiral worldsheet super-invariances: with $(4, 4)$ odd generators
in the $n=3$ case and $(8, 8)$ odd generators in the $n=5$ case. We also study closure of these supersymmetry
transformations, discuss the peculiarities of their on-shell realization and present the expression
for the relevant conserved supercurrent. Some concluding remarks are collected in Sect.~4.

\setcounter{equation}{0}
\section{Outline of Pohlmeyer reduction of the $AdS_n\times S^n$ \break superstring sigma models}

In this Section, following refs. \cite{TG, GT} \footnote{The systematic application
of the Pohlmeyer reduction to the bosonic coset models with demonstrating the equivalence of these PR systems
to certain mass-deformed gWZW models was earlier performed in \cite{B-O,BHS}.}, we briefly recall the main points
of the Pohlmeyer reduction procedure applied to $AdS_n\times S^n$ superstrings with $n=2,3,5\,$.

\subsection{Supercosets}
Superstring theories in a formulation with  manifest space-time supersymmetry
are naturally described as WZW-type sigma models with a supercoset target space.

For example, ${\cal N}=2$ Green-Schwarz superstring in $D=10$ Minkowski background can be
formulated as ${\cal P}/{\cal L}$ supercoset sigma model \cite{HM},
where ${\cal P}$ is ${\cal N}=2$, $D=10$ Poincar\'e supergroup,
${\cal L}$ is its Lorentz subgroup. The coset ${\cal P}/{\cal L}$ is just ${\cal N}=2, D=10$ Minkowski superspace.
This construction can be generalized to curved superbackgrounds,
in particular, to super $AdS_5\times S^5$ \cite{MT}. Besides this maximally supersymmetric $D=10$
background, one can consider non-critical $AdS$ string models in dimensions less than
$D=10$, namely on the superbackgrounds with the bosonic bodies $AdS_n\times S^n$ for $n<5$.
In all cases the superstring model is defined as $\hat F/G$ supercoset
sigma model, with $\hat F/G$ being an extension of the bosonic coset,
representing target space-time, to the corresponding supercoset. Namely,
the minimal superextensions of the target space-times
\bea
&& AdS_2\times S^2=\frac{SU(1,1)\times SU(2)}{U(1)\times U(1)}\,,\\
&& AdS_3\times S^3=\frac{SU(1,1)\times SU(1,1) \times SU(2)\times SU(2)} {SU(1,1)\times SU(2)}\,,\\
&& AdS_5\times S^5=\frac{SU(2,2)\times SU(4)}{SO(1,4)\times SO(5)}
\eea
are the following supercosets\footnote{Recall that the supergroup $PSU(m|m)$ is a quotient of $SU(m|m)$ over
the decoupling $U(1)$ generator and so has $2m^2 - 2$ bosonic parameters. }
\bea
&& AdS_2\times S^2:\quad\quad\frac{\hat F}{G}=\frac{PSU(1,1|2)}{U(1)
\times U(1)}\,,\\
&& AdS_3\times S^3:\quad\quad\frac{\hat F}{G}=\frac{PSU(1,1|2)\times
PSU(1,1|2)}{SU(1,1)\times SU(2)}\,,\\
&& AdS_5\times S^5:\quad\quad\frac{\hat F}{G}=\frac{PSU(2,2|4)}{SO(1,4)
\times SO(5)}\,.
\eea

\subsection{Constraints and gauge fixings}

The common feature of all three cases $n= 2, 3, 5$ is that the superalgebra $\hat f$ of the corresponding
supergroup $\hat F$ (or its complex special linear version) admits a $Z_4$ grading:
\be
\hat f=\hat f_0\oplus\hat f_1\oplus\hat f_2\oplus\hat f_3\label{Z4-dec}
\ee
with
\be
[\hat f_i,\hat f_j]\subset\hat f_{i+j\;mod\,4}\,.\label{Z4-comm}
\ee

Here $\hat f_0$ is the algebra of the bosonic group $G$, $\hat f_2$ is its orthogonal
complement to the full bosonic subalgebra of $\hat f\,$ and the subsets $\hat f_{1,3}$ are fermionic.
The currents $J_\pm=F^{-1}
\partial _\pm F$, where $F\in\hat F$, may be decomposed, according to
\p{Z4-dec}, as
\be
J_\pm={\cal A}_\pm+P_\pm+Q_{1\pm}+Q_{2\pm}\,,\quad\quad {\cal A}\in
\hat f_0\,,\quad Q_1\in\hat f_1\,,\quad P\in\hat f_2\,,\quad Q_2\in\hat f_3\,.
\ee

Lagrangian of the $\hat F/G$ superstring sigma model in the conformal gauge reads
\be
L_{GS}=\text{STr}[P_+P_-+\frac{1}{2}(Q_{1+}Q_{2-}-Q_{1-}Q_{2+})]\,.
\label{GS-L}
\ee
Here $\text{STr}$ denotes the  supertrace of supermatrix.
This formula is valid for any $n$, with further specializing of the contents of
the $P,\;Q$ currents in each particular case. In such a form the model involves
a number of redundant degrees of freedom, both bosonic and fermionic.
One way to eliminate them is the PR procedure combined with $\kappa$-symmetry gauge fixing.
It explicitly solves Virasoro constraints  and also keeps manifest $2d$ Lorentz symmetry.
To be more precise, the Lagrangian \p{GS-L} should be accompanied by Virasoro
constraints
\be
\text{STr}(P_+P_+)=0\,,\quad\text{STr}(P_-P_-)=0\,.\label{Virasoro}
\ee
The GS action is also invariant under fermionic $\kappa$-symmetry.
It can be partially fixed by setting
\be
Q_{1-}=0\,,\quad\quad Q_{2+}=0\,.
\ee

One can also partially fix the $\hat f_0$ gauge symmetry of \p{GS-L} and make use of the first
Virasoro constraint to set $P_+=p_+T$, where
$p_+=p_+(\sigma)$ and $T$ is a fixed element of $\hat f_2$. Usually $T$
is taken to be block-diagonal with the blocks $\frac{i}{2}\Sigma$,
where the matrix $\Sigma$ is used for hermitian conjugation of matrices with
Minkowski signature in the fundamental representation
space\footnote{In the $n=3$ case $\Sigma =\text{diag}\{1,-1\}$
and in the $n=5$ case $\Sigma=\text{diag}\{1,1,-1,-1\}\,$.}. The matrix $T$
allows one to split the superalgebra $\hat f$ as
\be
\hat f=\hat f^{||}\oplus\hat f^\bot\,,\quad P^{||}\zeta ^{||}=\zeta ^{||}\,,\quad P^{||}
\chi ^\bot =0\,,
\ee
where
\be
\zeta ^{||}\in\hat f^{||},\quad\chi ^\bot\in\hat f^\bot\,,\quad P^{||}=
-[T,\,[T,\,\cdot\,]]\,,\label{par-def}
\ee
and
\be
[\hat f^\bot,\,\hat f^\bot]\subset\hat f^\bot\,,\quad
[\hat f^{||},\,\hat f^\bot]\subset\hat f^{||}\,,\quad
[\hat f^{||},\,\hat f^{||}]\subset\hat f^\bot\,.\label{par-comm}
\ee
In particular, \p{par-def} implies that $T\in \hat{f}_2^{\bot}\,$. In what follows, we shall use some generic
properties of the matrix
$T$
\be
[T, f^\bot] = 0\,, \quad \{T, f^{||}\} = 0\,, \quad T^2 =
-\frac{1}{4}\,I\,.
\lb{Tpropert}
\ee

Using residual conformal invariance, one may fix $p^+=\mu$, where $\mu$ is some constant
with the mass dimension. Then the equation of motion $\partial _+P_-+[{\cal A_+},P_-]=0$ and the second Virasoro
constraint in \p{Virasoro} can be solved by setting $P_-=\mu g^{-1}Tg$, where
$g$ is a $G$-valued field. Finally, one uses the residual
$\kappa$-symmetry to entirely remove the non-physical fermionic degrees of freedom. The
remaining dynamical fermionic degrees of freedom are represented by the fields
\be
\Psi _R=\frac{1}{\sqrt{\mu}}Q_{1+}^{||} \in \hat{f}_1^{||}\,,\quad\quad\Psi _L=\frac{1}{\sqrt{\mu}}
(gQ_{2-}g^{-1})^{||} \in \hat{f}_3^{||}\,.\label{ferm-proj}
\ee

\subsection{Reduced equations and gWZW action}
The generalized Pohlmeyer reduction applied to the equations of motion associated with the Lagrangian \p{GS-L} finally
results in the following equations of motion for the reduced fields $g,\,\Psi _{L,R}$ (see details in \cite{TG, GT}):
\be
D_-(g^{-1}D_+g)-F_{+-}  = \mu ^2[T,\, g^{-1}Tg]+ \mu [\Psi_R, \, g^{-1}\Psi _Lg ]\,,
\label{pr-eq-1-1}
\ee
\be
D_-\Psi _R=\mu[T,\,g^{-1}\Psi _Lg]\,,\quad\quad
D_+\Psi _L=\mu[T,\,g\Psi _Rg^{-1}]\,,\label{pr-eq-22}
\ee
where the covariant derivatives are defined as $D_\pm=\partial _\pm+[A_\pm,\,\cdot\,]$ and
gauge field strength is
\be
F_{+-} = \partial_+A_- - \partial_-A_+ + [A_+,\,A_-]\,.
\ee

The $2d$ gauge fields $A_\pm$ take values in the algebra ${\tt h}$ of subgroup $H$ of
group $G$, defined by the condition
\be
[T,\,h]=0\,, \quad h\in{\tt h}\,,
\ee
so that
\be
{\tt g}=\hat f_0=  {\tt m}\oplus {\tt h}\,, \quad {\tt m} := \hat f_0^{||}\,, \;\; {\tt h} := \hat f_0^{\bot}\,.
\ee
In the  $n=2$ case, with $G=U(1)\times U(1)$,  the subgroup $H$ is empty and $A_\pm=0$. In the $n=3$ case,
with $G= SU(1,1)\times SU(2)$, we have $H =U(1)\times U(1)\,$, and in the $n=5$ case, with
$G=SO(1,4)\times SO(5)$, we have $H = SO(4) \times SO(4) \sim [SU(2)]^4\,$. Eqs. \p{pr-eq-1-1}, \p{pr-eq-22}
are covariant under the $H\times H$-valued gauge transformations
\be
g\rightarrow hg\bar h^{-1}\,,\quad \Psi _L\rightarrow h\Psi _Lh^{-1}\,,
\quad \Psi _R\rightarrow\bar h\Psi _R\bar h^{-1}\,,\nn
\ee
\be
A_+\rightarrow h(A_+ +\partial _+)h^{-1}\,,\quad
A_-\rightarrow\bar h(A_- +\partial _-)\bar h^{-1}\,.\label{act-gg}
\ee

It was found in \cite{TG,GT} that the equations  of motion \p{pr-eq-1-1}, \p{pr-eq-22}
are derivable from the following action
\be
S_{tot}=S_{gWZW}+\mu ^2\int d^2\sigma\text{STr}(g^{-1}TgT) \nn
\ee
\be
+\int d^2\sigma\left[\text{STr}(\Psi _LTD_+\Psi _L+\Psi _RTD_-\Psi _R)+\mu
\text{STr}(g\Psi _Rg^{-1}\Psi _L)\right],
\label{PR-Lagr-gen}
\ee
where $S_{gWZW}$ is the action of the $G/H$ gWZW model:
\bea
&& S_{gWZW} = S_{WZW} + S_{gauge}\,, \nn
&& S_{WZW}=\frac{1}{2}\int d^2\sigma\text{STr}\left(g^{-1}\partial _+gg^{-1}\partial _-g
\right)-\frac{1}{6}\int d^3\sigma\varepsilon ^{abc}\text{STr}\left(g^{-1}\partial _agg^{-1}
\partial _bgg^{-1}\partial _cg\right),\lb{WZW} \\
&& S_{gauge}=\int d^2\sigma\text{STr}(A_+\partial _-gg^{-1}-A_-g^{-1}\partial _+g-g^{-1}
A_+gA_-+A_+A_-)\,. \lb{gauge1}
\eea
In \p{WZW}, the second integral (WZ term) is taken over a three-dimensional space with the boundary
identified with the $2d$ base manifold\footnote{For brevity, and following \cite{GT},
we suppress the factors $\pi$ normally appearing
in the denominators of the coefficients.}.

The action \p{PR-Lagr-gen} is invariant under the $H$-valued gauge transformations
\be
g\rightarrow {h}g{h}{}^{-1}\,,\quad \Psi _{L,R}\rightarrow {h}\Psi _{L,R}{h}{}^{-1}\,,\quad
A_\pm\rightarrow {h}(A_\pm +\partial _\pm){h}{}^{-1}\,,\label{act-g}
\ee
which form a diagonal $h=\bar h$ in the ``on-shell'' gauge group $H\times H$ \p{act-gg}.

As a consequence of this Lagrangian formulation of the PR superstring equations \p{pr-eq-1-1}, \p{pr-eq-22},
there also appear additional algebraic constraints following from \p{PR-Lagr-gen} as equations of motion
for the gauge fields $A_\pm$:
\be
\mbox{\bf a)} \;\;(g^{-1}D_+g)_{\tt h}= 2(T\Psi _R^2)_{\tt h}\,, \quad  \mbox{\bf b)} \;\;
(g D_-g^{-1})_{\tt h}= 2(T\Psi _L^2)_{\tt h}\,.\label{A-eq}
\ee
As given in \cite{TG}, these equations can be interpreted as fixing of a certain gauge with respect
to the extended on-shell gauge group \p{act-gg}.
Also note that eqs. \p{A-eq}, being combined with eqs. \p{pr-eq-1-1}, \p{pr-eq-22}, imply that the
$2d$ gauge field strength vanishes on-shell:
\be
F_{+-}= 0\,. \lb{F-van}
\ee

Finally, we remark that the bosonic sector of the reduced model is described by the gWZW model on the coset
\be
\frac{SO(1,n-1)\times SO(n)}{SO(n-1)\times SO(n-1)}\,, \nn
\ee
with the action being a sum of $S_{gWZW}$ and the potential term $\sim \mu^2$ in \p{PR-Lagr-gen}.
This system is known to arise as a result of the Pohlmeyer reduction  applied to the bosonic string on
$AdS_n\times S^n$. The potential term is of the matrix sine-sinh-Gordon type.\\

\subsection{Worldsheet supersymmetry?}
The numbers of bosonic and fermionic degrees of freedom in the action \p{PR-Lagr-gen} properly match each other to suggest
the presence of hidden worldsheet supersymmetry in this fermionic extension of the gWZW action.
Indeed, it was shown in \cite{TG} that in the $n=2$ case the action possesses ${\cal N}=(2,2)$ supersymmetry
and is equivalent to the action of ${\cal N}=(2,2)$ sine-sinh-Gordon model \cite{sSG-1, sSG-2}. The relevant ${\cal N}=(2,0)$
supersymmetry transformations of this model can be formally generalized  to other $n$ as
\bea
&& \delta_{\epsilon_L}\, g = g[T, [\Psi_R, \,\epsilon_L]]\,, \quad \delta_{\epsilon_L}\,\Psi_R
= [(g^{-1}D_+ g)^{||}, \,\epsilon_L]\,, \quad \delta_{\epsilon_L}\,\Psi_L = \mu[T,\, g\epsilon_Lg^{-1}]\,, \nn
&& \delta_{\epsilon_L}\,A_+ = 0\,, \qquad \delta_{\epsilon_L}\,A_- = \mu [(g^{-1}\Psi _L g)^{\bot}, \,\epsilon_L]\,,
\label{susy-1}
\eea
where $\epsilon_L \in \hat{f}_1^\bot$ (and analogously for the right-handed supersymmetry,
with the parameter $\epsilon_R\in\hat f_3^\bot\,$). However, the action \p{PR-Lagr-gen} is invariant only
under the stringent condition $[\epsilon_L, \,h] =0$, which can be fulfilled only for $n=2$ \cite{TG}.
So far, no way was found to evade this obstruction against the off-shell worldsheet supersymmetry in the cases $n=3$ and $n=5\,$.

\setcounter{equation}{0}
\section{Modified mass-deformed gWZW action and its \break supersymmetry}

As a possible way of solving the problem of off-shell worldsheet
supersymmetry in the cases $n=3$ and $n=5$, we propose to derive the
PR form of superstring equations \p{pr-eq-1-1}, \p{pr-eq-22} from
some modification of the action \p{PR-Lagr-gen}, such that it includes a modified gWZW action.
While giving rise to the same PR superstring field equations \p{pr-eq-1-1}, \p{pr-eq-22}, it surprisingly
possesses a hidden ${\cal N}=(4,4)$ supersymmetry in the $n=3$ model and ${\cal N}=(8,8)$
supersymmetry in the $n=5$ model.

\subsection{An alternative action}
For what follows, it will prove important to systematically use the Polyakov-Wiegmann representation
for the gauge fields $A_\pm$:
\be
A_+=-\partial _+uu^{-1}\,,\quad\quad
A_-=-\partial _-\bar u\bar u^{-1}\,,\label{Acur}
\ee
where $u$ and $\bar u$ are two independent matrices with values in the group $H$.
The general $H\times H$ gauge transformation laws of $A_\pm$ (see \p{act-gg}) are reproduced by the following gauge transformation
laws of the ``prepotentials'' $u$ and $\bar u$:
\be
u\rightarrow hu\,,\quad\quad\bar u\rightarrow\bar h\bar u\,. \lb{ubaru2}
\ee
Note that the definition \p{Acur} is not changed under the additional right gauge transformations of $u, \bar u$ with holomorphic and
anti-holomorphic parameters (Kac-Moody (KM) symmetries)
\be
u \rightarrow u\, \omega(\sigma^-)\,,\quad\quad \bar u\rightarrow \bar u\, \bar\omega(\sigma^+)\,. \lb{holtran}
\ee

The representation \p{Acur} is well known and, in the PR superstring context, was already used in \cite{TG} and \cite{RT}
for different purposes. For instance, in the paper \cite{RT} devoted to analyzing the UV-finiteness properties of the
PR $AdS_5\times S^5$ superstring theory, the gauge fields
in the action  \p{PR-Lagr-gen} were substituted as in \p{Acur} to isolate gauge degrees of freedom via the Polyakov-Wiegmann
identity for $S_{gWZW}$. We propose to take \p{Acur} as an input and to consider just $u$ and $\bar u$ as the basic gauge objects,
both at the classical and quantum levels. As an important consequence, the equations of motion for the gauge fields
will be of the second order in derivatives, as distinct from the algebraic equations \p{A-eq} of the standard approach.
Also, as we shall see,
treating $u$ and $\bar u$ as the basic entities provides some additional possibilities for implementing new symmetries in the action.

Let us turn to our main point. We propose the following modified action for the PR superstrings:
\be
S_{tot}\rightarrow S'_{tot} = S_{tot}+S_a\,,\label{mod-act}
\ee
where $S_{tot}$ is the standard action \p{PR-Lagr-gen} (with the gauge fields represented according to \p{Acur}) and
\be
S_a=S_{WZW}^{(H)}(B) \lb{newSa}
\ee
is WZW action for the $H$-valued field $B=u^{-1}\bar u$. The field
$B$ is manifestly invariant under the diagonal $h=\bar h$ subgroup of the gauge transformations \p{ubaru2}, so the addition \p{newSa}
and the new total action $S_{tot}'$ are also gauge-invariant.

The equations of motion for the fields $g\,,\Psi _{L,R}$, \p{pr-eq-1-1} and \p{pr-eq-22},  {\it are not changed}
upon the modification \p{mod-act}, it affects only equations of motion for the fields
$u,\,\bar u\,$. This is important because the Pohlmeyer-reduction approach, in its own right,
gives rise just to the equations \p{pr-eq-1-1} and \p{pr-eq-22}.
The appearance of additional equations for the gauge fields is a ``price'' for the possibility
to derive the equations \p{pr-eq-1-1}, \p{pr-eq-22} from an off-shell action,
and, for the action \p{PR-Lagr-gen}, these equations (i.e. \p{A-eq}) may be regarded as a partial fixing
of the $H\times H$ gauge freedom of the PR superstring equations \p{pr-eq-1-1}, \p{pr-eq-22}.

The equations of motion for gauge fields corresponding to the action $S'_{tot}$ are now different from \p{A-eq}.
Using the general formula for the variation of $S_a$:
\be
\delta S_a=-\int d^2\sigma\text{STr}(B^{-1}\delta B\partial _-(B^{-1}\partial _+B))=
\int d^2\sigma \text{STr}\left(\delta\bar u\bar u^{-1}\,F_{+-} - \delta u u^{-1}\,F_{+-}\right), \lb{varSa}
\ee
and the properties
\be
\delta A_+ = -D_+(\delta u u^{-1})\,,\quad\quad\delta A_-=-D_-(\delta\bar u\bar u^{-1})\,,\label{dAAn}
\ee
we derive
\be
\mbox{\bf a)} \;\;D_-\left(g^{-1}D_+g-2T\Psi _R^2\right)_{\tt h}-F_{+-}=0\,, \quad \mbox{\bf b)} \;\;
D_+\left(D_-g g^{-1} +2T\Psi _L^2\right)_{\tt h}-F_{+-}=0\,. \lb{A-eqmod}
\ee

Using the fermionic equations  \p{pr-eq-22} in \p{A-eqmod} and comparing the result with eq. \p{pr-eq-1-1}, we observe that
eqs. \p{A-eqmod} are just two equivalent forms of the ${\tt h}$ projection of \p{pr-eq-1-1}. Thus in the present case
the gauge field equations are identically satisfied as a consequence of the PR superstring equations \p{pr-eq-1-1}, \p{pr-eq-22}.
No any constraint on the gauge fields appear. Note that at any other coefficient before $S_a$ in \p{mod-act} these properties
would be lost, though the equations of motion for $g, \Psi_L, \Psi_R$ would be the same\footnote{In this case, e.g.,
the equations for the gauge fields would imply $F_{+ -} =0\,$.}. As we shall see, the hidden supersymmetry of $S'_{tot}$
is also revealed only at this special value of the coefficient.

The reason why the combination
\be
S'_{gWZW} = S_{gWZW} + S_a
\ee
is distinguished among other linear combinations of these two actions becomes clear after
using the well known consequence of Polyakov-Wiegmann identity \cite{Pol-Wig-2}
\be
S_{gWZW}(g, u, \bar u) = S_{WZW}(u^{-1} g \bar u) - S_{WZW}^{(H)}(u^{-1} \bar u) \,,
\ee
whence
\be
S'_{gWZW}(g, u, \bar u) = S_{WZW}(u^{-1} g \bar u)\,.\lb{PWform}
\ee

Using \p{PWform} and the property that $[T, h] =0$, it is easy to show that the action $S'_{tot}$ is invariant not only
under the diagonal gauge $H$ subgroup \p{act-g} (as occurs for $S_{tot}$), but under the full gauge $H\times H$ group \p{act-gg}
which now defines an {\it off-shell} gauge invariance\footnote{This extended gauge symmetry is of course reduced
to its diagonal subgroup for any other coefficient in front of $S_a$ in \p{mod-act}.}. The new action is also invariant
under the (anti)holomorphic right KM shifts \p{holtran}.

For further use, we define the ``shadow'' gauge fields $\tilde{A}_\pm$:
\be
\tilde{A}_+ = -\partial _+ \bar u \bar u^{-1}\,, \quad \tilde{A}_- = -\partial _-u  u^{-1}\,.\label{Acur1}
\ee
They satisfy the mixed flatness conditions
\bea
&& \tilde{F}_{+ -} = \partial_+ \tilde{A}_- -\partial_-{A}_+  + [A_+, \tilde{A}_-] = D_+\tilde{A}_- - \partial_-{A}_+  = 0\,, \nn
&& \tilde{F}_{- +} = \partial_- \tilde{A}_+ -\partial_+ {A}_-  + [A_-, \tilde{A}_+] = D_-\tilde{A}_+ - \partial_+{A}_-  = 0\,, \lb{shadF}
\eea
and have ``twisted'' transformation laws under the gauge $H\times H$ group \p{ubaru2}:
\be
\tilde{A}_+ \rightarrow \bar h(\tilde{A}_+ +\partial _+)\bar h^{-1}\,,\quad
\tilde{A}_-\rightarrow h(\tilde{A}_- +\partial _-) h^{-1}\,.\label{act-ggShad}
\ee
The conditions \p{shadF} are valid off shell and can in fact serve as the definition of the shadow gauge fields. Note
that $\tilde{A}_\pm$ are transformed under the (anti)holomorphic KM transformations \p{holtran}, but the mixed
field strengths $\tilde{F}_{+ -}, \tilde{F}_{- +}$ are invariant with respect to them due to the properties
$D_\pm \tilde{A}_\mp' = D_\pm \tilde{A}_\mp\,$.

For what follows we shall need some convenient expressions for the full variations of the action $S'_{tot}$ with respect to
general $\delta u$ and $\delta \bar u\,$:
\bea
&& \delta_u S'_{tot} = \int d^2\sigma\,\text{STr}\left[\delta u u^{-1}\,D_+\left(\partial_-g g^{-1} - gA_-g^{-1} + \tilde{A}_-
+ 2T \Psi_L^2\right)_{\tt h}\right], \label{Avar-11} \\
&& \delta_{\bar u} S'_{tot} = -\int d^2\sigma\,\text{STr}\left[\delta \bar u \bar u^{-1}\,
D_-\left(g^{-1} \partial_+g +g^{-1 }A_+g- \tilde{A}_+ - 2T \Psi_R^2\right)_{\tt h}\right].\label{Avar-1}
\eea
The variations of the additional term $S_a$ can also be cast in a similar form:
\be
\delta S_a= \int d^2\sigma \text{STr}\left[\delta u u^{-1}\,D_+(\tilde{A}_- - {A}_-)+
\delta \bar u \bar u^{-1}\,D_-(\tilde{A}_+ - {A}_+)\right].
\lb{varSa2}
\ee

\subsection{Off-shell worldsheet supersymmetry}

As a prototype for the off-shell supersymmetry transformations of the PR superstring action we take the transformations
\p{susy-1}. As was already mentioned, they provide formal symmetries of the action \p{PR-Lagr-gen}
if the condition $[\epsilon _L,\,h]=0$ is satisfied, and the main problem consists in that this condition is too strong, being
achievable only for $n=2\,$. It turns out that, within the setting described in the previous subsection,
we can give up this restriction by modifying \p{susy-1} as
\bea
&& \delta_{\epsilon_L}\, g = g[T, [\Psi_R, \,\tilde\epsilon_L]]\,, \quad \delta_{\epsilon_L}\,\Psi_R
= [(g^{-1}D_+ g)^{||}, \,\tilde\epsilon_L]\,, \quad \delta_{\epsilon_L}\,\Psi_L = \mu[T,\, g\tilde\epsilon_Lg^{-1}]\,, \label{susy-3} \\
&& \delta_{\epsilon_L}\,A_+ = 0\,, \qquad \delta_{\epsilon_L}\,A_- = \mu [(g^{-1}\Psi _L g)^{\bot}, \,\tilde\epsilon_L]\,,
\label{susy-2}
\eea
where\footnote{The matrix $\epsilon _L\in\hat f_1^\bot$ encompasses $2(n-1)$ independent
parameters for the $AdS_n\times S^n$ model. The transformations of the right chiral supersymmetry can be written
in a symmetric way through the matrix parameter $\tilde{\epsilon}_R = u\epsilon_R u^{-1}$ with the same number of independent entries.}
\be
\tilde\epsilon _L=\bar u\epsilon _L\bar u^{-1}\,.\label{te}
\ee

To show that the action $S_{tot}'$ is indeed invariant, we start with the massless $\mu = 0$ case. In this case, the
variation of $S_{tot}'$ coincides with that of $S_{tot}$, because the gauge fields and, hence, the addition $S_a$, are not varied.
The key role in checking the invariance is played by the relation
\be
D_-\tilde\epsilon _L=0\,,
\ee
and no need in imposing the additional requirement $[\epsilon _L,\,h]=0$ arises.

As the second step, let us vary the massive action $S_{tot}$, still assuming that $\delta A_- = 0\,$.
Once again, this variation is equal to that of the modified action $S_{tot}'$ and, up to a total derivative under the $2d$ integral,
is found to be:
\be
\mu\int d^2\sigma\text{STr}\left(g^{-1}\Psi _Lg\left[\tilde\epsilon _L,\,(g^{-1}\partial _+g+g^{-1}A_+g-\tilde A_+-2T\Psi _R^2)_{\tt h}\right]
\right).\label{masvar}
\ee
Then, recalling the general formula \p{Avar-1} for variations of the full modified action $S_{tot}'$ with respect to $\delta \bar u$
and the property that $\delta A_- = -D_-(\delta \bar u \bar u^{-1})$ (see \p{dAAn}), we find that the variation \p{masvar}
is exactly canceled by the contribution coming from the variation of $A_-$ according to \p{susy-2}.

The crucial role in this cancelation is played by the presence of the additional piece $S_a$ in $S_{tot}'$ as compared to $S_{tot}\,$.
Without this term, the $\delta A_-$ variation is
\be
\delta (S'_{tot} - S_a) = -\mu \int d^2\sigma\,\text{STr}\left(g^{-1}\Psi _Lg\left[\tilde\epsilon _L,\,
(g^{-1}\partial _+g+g^{-1}A_+g- A_+ -2T\Psi _R^2)_{\tt h}\right] \right).\label{Avar-2}
\ee
It only partly cancels \p{masvar}, and for the vanishing of the total variation one is led to require $[\epsilon _L\,,h]=0\,.$
No such a restriction is necessary if the term $S_a$ is added.

It is worthwhile to note that the local transformation \p{susy-2} of the gauge potential $A_-$ amounts
to a non-local transformation of the prepotential $\bar u$. It is obtained as a solution of the equation
\be
D_-(\delta\bar u\bar u^{-1})=\mu[\tilde\epsilon _L,\,(g^{-1}\Psi _Lg)^\bot] \quad \Rightarrow \quad
\bar u^{-1}\delta\bar u = \mu\,(\partial _-)^{-1}\,\left(\bar u^{-1}[\tilde\epsilon _L,\,(g^{-1}\Psi _Lg)^\bot]\bar u\right).\label{invu}
\ee
Some zero-mode holomorphic ${\tt h}$-valued  function $f(\sigma^+)$ arising as an integration constant of the solution \p{invu},
can be absorbed into the KM-type transformation \p{holtran} of $\bar u$.

The consideration in this Subsection is generic for both $n=3$ and $n=5$ cases. The difference from the $n=2$ case
is that now there are gauge fields and prepotentials, non-trivial transformations of which ensure the invariance
of the action. In principle, the matrix prepotential $\bar u\,$, which appears in \p{susy-3}, \p{susy-2} through the dressing relation \p{te},
can be non-locally expressed in terms of $A_-$ (up to holomorphic right KM transformation):
$$
(\partial_- + A_-)\bar u = 0 \quad \Rightarrow \quad \bar u = P\exp\{-\int^{\sigma^-}d\sigma^-{}' A_-(\sigma^-{}', \sigma^+)\}\,
\bar\omega(\sigma^+)\,.
$$
The same concerns the prepotential $u$ which is present in the right-handed  chiral
supersymmetry transformations: it can be expressed through the gauge field $A_+$. So the transformations \p{susy-3}, \p{susy-2}
and \p{invu}, as well as their right-handed counterparts, can be entirely expressed in terms of the objects actually entering
the PR superstring equations \p{pr-eq-1-1} and \p{pr-eq-22}, i.e. in terms of $2d$ fields $g, \Psi_L, \Psi_R$ and $A_{\pm}\,$.

Note that these transformations are simplified in the gauge
\be
\bar u = u = I\,. \lb{Gau2}
\ee
In this gauge, $A_\pm = 0$ and all supersymmetry transformations acquire extra terms corresponding to
the compensating $H\times H$ gauge transformations needed to preserve \p{Gau2}. In particular, the $\epsilon_L$ transformations become
\bea
&& \delta_{\epsilon_L}\, g = g([T, [\Psi_R, \,\epsilon_L]] + \hat\delta \bar h)\,, \quad \delta_{\epsilon_L}\,\Psi_R
= [(g^{-1}\partial _+ g)^{||}, \,\epsilon_L] + [\Psi_R, \,\hat\delta \bar h]\,, \nn
&& \delta_{\epsilon_L}\,\Psi_L = \mu[T,\, g\epsilon_Lg^{-1}]\,, \quad \delta A_\pm = 0\,, \lb{tranGau2}
\eea
where $\hat\delta \bar h = \mu (\partial_-)^{-1}\,[\epsilon_L, (g^{-1}\Psi_Lg)^\bot]\,$. Thus the non-locality remains
in the gauge \p{Gau2} too.

\subsection{On-shell supersymmetry}
Let us now study how the off-shell supersymmetry of $S_{tot}'$ is implemented on the corresponding equations of motion.
As we know, these are just the PR superstring equations \p{pr-eq-1-1} and \p{pr-eq-22}.

It is rather straightforward to check that, under the transformations \p{susy-3} and \p{susy-2}, these equations are transformed as follows
\bea
&& \delta\left(D_+\Psi _L-\mu [T,\,g\Psi _Rg^{-1}]\right)=2\mu T\left(g[{\cal O}_+,\, \tilde\epsilon]g^{-1}\right)^{||}\,,\nn
&& \delta\left(D_-\Psi _R-\mu [T,\,g^{-1}\Psi _Lg]\right)=0\,,\nn
&& \delta\left(D_-(g^{-1}D_+g)-F_{+-}-\mu ^2[T,\,g^{-1}Tg]-\mu [\Psi _R,\,g^{-1}\Psi _Lg]\right)=
\mu [g^{-1}\Psi _Lg,\,[\tilde\epsilon ,\, {\cal O}_+]],\lb{VarUr}
\eea
where
\be
{\cal O}_+ = (g^{-1}\partial _+g+g^{-1}A_+g-2T\Psi _R^2-\tilde A_+)_{\tt h}\,.\lb{defO}
\ee

We observe an interesting deviation from what one could expect by analogy with the standard supersymmetric theories:
whereas the action $S_{tot}'$ is invariant under \p{susy-3}, \p{susy-2} and \p{invu}, the equations of motion are not, they involve
a non-vanishing object ${\cal O}_+$ in their right-hand sides. This can be related to the non-standard fact
that the fundamental entities of the action, the prepotentials $\bar u$ (or $u$ in the case of the right-handed supersymmetry),
undergo the non-local transformation \p{invu}.

Nevertheless, it turns out that the equations of motion can be made invariant at cost of slight modification
of the transformations \p{susy-3}, \p{susy-2} and \p{invu} on shell. First we notice the relation
\be
D_-{\cal O}_+ = 0\,,
\ee
which is satisfied as the ${\tt h}$ projection of the bosonic equation \p{pr-eq-1-1}, with taking into account the fermionic equation \p{pr-eq-22}.
Then the current $\tilde{{\cal O}}_+ = \bar u^{-1}{\cal O}_+ \bar u$
satisfies the conservation law
\be
\partial_-\tilde{{\cal O}}_+ = 0 \; \Rightarrow \; \tilde{{\cal O}}_+ = \lambda(\sigma^+)\,, \;{\cal O}_+
= \bar u \, \lambda(\sigma^+)\, \bar u^{-1}\,.
\ee
Based on this representation, one can re-express  $\tilde{{\cal O}}_+$ on shell through the holomorphic $H$-valued matrices
$\hat{\bar\omega} (\sigma^+)$ as
\be
\tilde{{\cal O}}_+ = \hat{\bar\omega}\,\partial_+\,\hat{\bar\omega}{}^{-1}\;. \lb{Osolv}
\ee
{}From this relation, $\hat{\bar\omega}$ can be non-locally expressed through $\tilde{{\cal O}}_+$ and, hence, through the basic fields
$g, \Psi_R$ and $\bar u\,$. As the last step, we modify the transformations \p{susy-3}, \p{susy-2} and \p{invu} by replacing
\be
\epsilon_L \;\Rightarrow \; \hat{\bar\omega} \,\epsilon_L\, \hat{\bar\omega}{}^{-1}\,.
\ee
Then, using \p{Osolv}, it is easy to check that the whole set of eqs. \p{pr-eq-1-1} and \p{pr-eq-22} is invariant under
such modified on-shell transformations.

The current $\tilde{{\cal O}}_+$ is invariant under the gauge $H\times H$ transformations, but behaves as a gauge connection with
respect to the holomorphic KM transformations \p{holtran}:
\be
\tilde{{\cal O}}_+ \quad \rightarrow  \quad \bar\omega^{-1}\,(\tilde{{\cal O}}_+ + \partial_+) \bar\omega\,,
\ee
or, in terms of the on-shell prepotential $\hat{\bar\omega}$,
\be
\hat{\bar\omega} \quad \rightarrow  \quad \bar\omega^{-1}\,\hat{\bar\omega}\,.
\ee
Hence, one can choose the on-shell gauge
\be
\hat{\bar\omega} = I \quad \Leftrightarrow \quad \tilde{{\cal O}}_+ = {\cal O}_+ = 0\,. \lb{gauge}
\ee
It is easy to check that in this gauge the current $\tilde{{\cal O}}_+$ is transformed under supersymmetry as
\bea
\delta \tilde{{\cal O}}_+ = \bar u^{-1}\left(2T [\tilde{\epsilon}_L, \,\{(g^{-1}D_+ g)^{||}, \Psi_R \}]
+ \mu \tilde{D}_+ \left(\bar u\,(\partial_-)^{-1}\left(\bar u^{-1}[\tilde{\epsilon}_L, \,(g^{-1}\Psi_L g)^\bot]\bar u\right)\,
\bar u^{-1}\right) \right)\bar u\,,
\eea
where $\tilde{D}_+ =\partial_+ + [\tilde{A}_+, .]\,$. One can also check that $\partial_-\delta \tilde{{\cal O}}_+  = 0$
as a consequence of eqs. \p{pr-eq-1-1}, \p{pr-eq-22} and the gauge condition \p{gauge}. Then, to preserve the gauge \p{gauge},
one should accompany the supersymmetry transformations by some field-dependent KM transformation. Obviously, \p{pr-eq-1-1}
and \p{pr-eq-22} remain invariant under such modified transformations, since they are invariant under arbitrary KM transformations.

\subsection{Conserved supercurrent and on-shell degrees of freedom}
The characteristic feature of supersymmetric systems is the existence of conserved supercurrent by which the corresponding
Noether supercharges can be constructed.

To find it in the case under consideration, we apply the standard procedure: vary $S_{tot}'$ with respect
to the group variations \p{susy-3}, \p{susy-2}, \p{invu}, in which the substitution
$\epsilon_L \rightarrow \epsilon_L(\sigma^+, \sigma^-)$ was made. Then the components of the supercurrent can be found from
\be
\delta S_{tot}=\int d^2\sigma\text{STr}(\partial _+ \epsilon _LJ_-+\partial _- \epsilon _LJ_+)\,.
\ee
Explicitly,
\bea
&& J_+=\bar u^{-1}[(g^{-1}D_+g)^{||},\,[T,\,\Psi _R]]\bar u +
\mu[\partial _-^{-1}(\bar u^{-1}(g^{-1}\Psi _Lg)^\bot\bar u),{\tilde{\cal O}}_+]\,,\nn
&& J_-=-\mu\bar u^{-1}(g^{-1}\Psi _Lg)^\bot\bar u\,.
\eea
It is straightforward to check that, when equations of motion \p{pr-eq-1-1} and \p{pr-eq-22} are satisfied, the supercurrent obeys
the standard conservation law
\be
\partial _+J_-+\partial _-J_+=0\,.
\ee
It is worth noting that the non-local term in $J_+$ disappears in the on-shell gauge \p{gauge}. An analogous conserved supercurrent
can be defined for the $\epsilon_R$ supersymmetry.

We  finish this subsection by a few comments concerning the on-shell degrees of freedom. Since the equations of motion
following from the action $S_{tot}'$ are none other than the PR form of the $AdS_n\times S^n$ superstring equations of motion,
without any additional restrictions on the gauge fields, all arguments of ref. \cite{TG} are applicable to the present case too.
In particular, the gauge $H\times H$ freedom of these equations can be fixed to implement the constraints \p{A-eq}
on the gauge fields as a particular choice of gauge, with the diagonal subgroup $h=\bar h$ as the residual gauge symmetry. Then the latter
can be used to reduce the number of independent bosonic degrees of freedom in the matrix $g$ to $({\rm dim}\, G - {\rm dim}\,H)$
and \p{A-eq} can be used to eliminate the gauge fields $A_\pm$ in terms of the physical bosonic and fermionic fields. As a result,
on shell we are left with the ``bosonic + fermionic'' field contents $(2 + 2)$, $(4 + 4)$ and $(8 + 8)$ in the cases $n=2$, $n=3$ and $n=5$,
respectively.

The prepotential representation \p{Acur} for the gauge fields provides some equivalent ways to reach the same conclusions.
One can choose the gauge \p{Gau2}, which is attainable both on and off shell. Its residual gauge group consists of the
$H\times H$ gauge transformations of the special form
\be
h = \omega^{-1}(\sigma^-)\,, \quad \bar h = \bar\omega^{-1}(\sigma^+)\,,\lb{resid2}
\ee
where we made use of the fact that on the general $u$ and $\bar u$ both the $H\times H$ gauge transformations \p{ubaru2}
and the (anti)holomorphic KM transformations \p{holtran} are realized. In this gauge, the ${\tt h}$-projections of
eq. \p{pr-eq-1-1}, with taking account of eq. \p{pr-eq-22}, become
\be
\partial _-(g^{-1}\partial _+g-2T\Psi _R^2)_{\tt h}=0\,, \quad \partial _+(\partial _-gg^{-1}+2T\Psi _L^2)_{\tt h}=0\,.
\ee
Since the expressions within the round brackets do not depend, respectively, on $\sigma^-$ and $\sigma^+$, the residual gauge freedom \p{resid2}
allows one to impose the following on-shell gauges
\be
(g^{-1}\partial _+g-2T\Psi _R^2)_{\tt h}=0\,, \quad (\partial _-gg^{-1}+2T\Psi _L^2)_{\tt h}=0\,. \lb{KMgauge1}
\ee
As a result, the $H$-subgroup degrees of freedom in the field $g$ prove to be eliminated. There exists another form of the
on-shell gauge \p{KMgauge1}, such that it still reveals the manifest $H\times H$ gauge covariance:
\be
{\cal O}_+ = {\cal O}_- = 0\,, \lb{Ogauge}
\ee
where ${\cal O}_+$ was defined in \p{defO} and ${\cal O}_-$ is its right-handed counterpart:
\be
{\cal O}_- = \left(\partial_-g g^{-1} - gA_-g^{-1} + \tilde{A}_- + 2 T\Psi_L^2\right)_{\tt h}.
\ee
In the gauge \p{Gau2}, we recover \p{KMgauge1}.

\subsection{Closure}
Let us study the on-shell closure of supersymmetry transformations. First of all, we can exploit
the $H\times H$ gauge symmetry to choose the gauge \p{Gau2} and consider the fixed-gauge form of supersymmetry transformations \p{tranGau2}.
The direct calculation of their Lie bracket on the fields $g,\,\Psi _{L}$ and $\Psi_R$ gives,
up to some additional (gauge) transformation terms, the translation terms multiplied by the same unique bracket matrix  parameter:
\bea
&& (\delta _1\delta _2-\delta _2\delta _1)\,g=4\partial _+gT(\epsilon _{L(2)}\epsilon _{L(1)}
-\epsilon _{L(1)}\epsilon _{L(2)})+\cdots\,, \nn
&& (\delta _1\delta _2-\delta _2\delta _1)\,\Psi _R=4\partial _+\Psi _RT(\epsilon _{L(2)}\epsilon _{L(1)}-\epsilon _{L(1)}\epsilon _{L(2)})+\cdots\,,
\nn
&& (\delta _1\delta _2-\delta _2\delta _1)\,\Psi _L= 4\partial _+\Psi _LT(\epsilon _{L(2)}\epsilon _{L(1)}-\epsilon _{L(1)}\epsilon _{L(2)})
+\cdots\,.\label{psiLLie-1}
\eea

Now, our purpose is to single out the ``genuine'' translation term multiplied by some $c$-number bracket parameter. To this end,
we have to take into account the detailed structure of the matrix-valued supersymmetry parameter $\epsilon _L$.
We relax the strong gauge condition \p{Gau2} and will firstly proceed without any gauge-fixing at all. The form of $\epsilon_L$
is uniquely determined by the condition $\epsilon_{L} \in \hat{f}_1^\bot$ implying $[\epsilon_L, T] = 0\,$. We have
\be
\epsilon_L =\left({0\atop iE^\dagger \Sigma}\;{E\atop 0}\right),\label{epsgenpar}
\ee
and, consequently, obtain the following expression for the matrix bracket parameter:
\be
\epsilon _{L(2)}\epsilon _{L(1)}-\epsilon _{L(1)}\epsilon _{L(2)}=
2\left({E_{(2)}E_{(1)}^\dagger -E_{(1)}E_{(2)}^\dagger\atop 0}\;{0\atop E_{(2)}^\dagger E_{(1)}-E_{(1)}^\dagger E_{(2)}}\right)T\,.
\ee
\vspace{0.3cm}

\noindent Consider \underline{$n=3$ model}. The most general form of the matrix $E$ in \p{epsgenpar} is determined by the aforementioned
conditions on $\epsilon_L$ as follows
\be
E^1=\eta ^1I_2\,,\;\;\text{or}\;\;E^2=\eta ^2\Sigma\,,
\ee
where $\eta ^1$ and $\eta ^2$ are complex Grassmann parameters, and $I_2$ is the unit $2\times 2$ matrix.
Then, for each of $E^i\,, i=1,2\,$, we obtain
(no summation over $i$)
\be
E^i_{(2)}E_{(1)}^{i\dagger} -E^i_{(1)}E_{(2)}^{i\dagger} =E_{(2)}^{i\dagger} E^i_{(1)}-E_{(1)}^{i\dagger} E^i_{(2)}=a^+_iI_2\,,\label{EE-ee1}
\ee
where we have defined two bracket translation parameters as
\be
a_i^+=\eta  ^i_{(2)}\eta  _{(1)}^{i\dagger}-\eta  ^i_{(1)}\eta  _{(2)}^{i\dagger}\,.\label{ai}
\ee
\vspace{0.3cm}

\noindent Consider \underline{$n=5$ model}. In this case the matrix $E$ in \p{epsgenpar} can be parametrized as follows:
\be
E=\left({\tilde E\atop 0}\;{0\atop\tilde H}\right),\label{En=5'}
\ee
where
\be
\tilde E=\left({\eta \atop 0}\;{0\atop\eta  ^\dagger}\right),\;\;\text{or}\;\;
\left({0 \atop -\eta ^\dagger}\;{\eta\atop 0}\right),\quad
\tilde H=\left({\eta \atop 0}\;{0\atop -\eta  ^\dagger}\right),\;\;\text{or}\;\;
\left({0 \atop \eta ^\dagger}\;{\eta\atop 0}\right).
\ee
We can form four possible combinations of these blocks to construct the matrix $E$ in \p{En=5'}.
It is easy to see that in each case  we again obtain the relation \p{EE-ee1}, now for four different
complex Grassmann parameters of supersymmetry $\eta ^i\,$, $i=1,\ldots 4\,$. Correspondingly, we get four different translation
bracket parameters of the form \p{ai}.

As soon as the relations \p{EE-ee1} are satisfied, we obtain\footnote{Due to the relation $[T,\,{\tt h}]=0\,$,
we can make the replacement $\epsilon_L \;\rightarrow \;\tilde\epsilon _L$ in \p{EEa-cl} without appearance of field-dependent
terms in $a^+_i$. This should be taken into account when studying Lie brackets without imposing the gauge \p{Gau2}.}
\be
\epsilon _{L(2)}\epsilon _{L(1)}-\epsilon _{L(1)}\epsilon _{L(2)}=2 a^+_iT\,.\label{EEa-cl}
\ee
Coming back to the gauge \p{Gau2}, we observe that \p{EEa-cl} implies the presence of the standard $a_i^+$-translation
terms in the Lie brackets \p{psiLLie-1}, as the only terms which are linear in fields.

It is convenient to choose the on-shell gauge \p{Ogauge} instead of the very restrictive gauge \p{Gau2}.
Then for each pair of the $\eta ^i$-parametrized supersymmetry transformations we obtain the following exact $(4,0)$ ($(8,0)$)
Lie brackets between the $\eta$ and $\eta^\dagger$ transformations (for brevity, we suppress the  index $i$ of $a^+_i$):
\bea
&& (\delta _1\delta _2-\delta _2\delta _1)\,g=-2a^+\partial _+g-2a^+A_+\,g+g (2a^+ A_+ + \tilde Q)\,,\nn
&& (\delta _1\delta _2-\delta _2\delta _1)\,\Psi _R=-2a^+\partial _+\Psi _R+[\Psi _R,\,2a^+A_+ +\tilde Q]\,,\nn
&& (\delta _1\delta _2-\delta _2\delta _1)\,\Psi _L= -2a^+\partial _+\Psi _L- [2a^+A_+,\,\Psi _L]\,,\nn
&& (\delta _1\delta _2-\delta _2\delta _1)\,A_-= -2a^+\partial _+A_- + D_-(2a^+ A_++ \tilde Q)\,,\nn
&& (\delta _1\delta _2-\delta _2\delta _1)\,A_+= -2a^+\partial _+A_++ D_+(2a^+ A_+)=0\,.\label{A+LLie}
\eea
Here $\tilde Q=Q+2a^+(\tilde A_+-A_+)$ and $Q$ is some field-dependent ${\tt h}$-valued matrix. The structure
of the closure in the considered sector is nicely transparent: it is a sum of the $2d$ $\sigma^+$ translations and compensating
field-dependent $H\times H$ gauge transformations, with the parameters $-(2a^+ A_+)$ and $-(2a^+A_+ + \tilde Q)$, respectively.
It has the unique form for all involved fields, as should be. While calculating these brackets, we used, besides the on-shell
gauge \p{Ogauge}, also the equations of motion for $\Psi_{L,R}\,$. It is also of interest to calculate the on-shell Lie brackets between
different pairs of ${\cal N}=(2,2)$ transformations and between the transformations with the (anti)holomorphic parameters within each
pair.

This study, equally as the detailed analysis of the closure properties of the full supersymmetry algebra, both on and off shell,
will be performed elsewhere. The transformations in the closure should clearly be symmetries of the equations
of motion \p{pr-eq-1-1} and \p{pr-eq-22}, as well as of the action $S_{tot}'$, eq. \p{mod-act}. It is interesting
to see whether these symmetries are reduced to the known ones (i.e. $H\times H$ gauge symmetries and (anti)holomorphic KM symmetries),
or they also contain some extra hidden symmetries.

\setcounter{equation}{0}
\section{Conclusions}
In the present paper, using a modified off-shell gWZW-type action for the PR $AdS_n\times S^n$ superstring equations, we revealed the existence
of hidden ${\cal N}=(4,4)$ and ${\cal N}=(8,8)$ chiral supersymmetries in these systems for $n=3$ and $n=5$
cases\footnote{It was suggested in \cite{Schm} that the
PR $AdS_3\times S^3$ superstring could possess a hidden ${\cal N}=(4,4)$ supersymmetry.}. We gave the explicit form
of the corresponding transformations, both on the off-shell level of the action and on the on-shell level of the equations
of motion (i.e. the PR superstring equations). These transformations necessarily contain non-local terms which arise
as a result of using the Polyakov-Wiegmann representation for the $2d$ gauge fields in the gWZW action. Modulo possible
field-dependent ``central charges'' in the crossing Lie brackets,
the supersymmetries found contain two (in the $n=3$ case) and four (in the $n=5$ case) independent ${\cal N}=(2,2)$ Poincar\'e subgroups
with the standard on-shell closure on the worldsheet translations (accompanied by field-dependent compensating gauge transformations).

It still remains to clarify what kind of extended $2d$ supersymmetry, or its generalization, we are facing in these systems.
Obviously, in order to understand this we need a genuine off-shell formulation of these models in terms of the appropriate
off-shell supermultiplets, with equal numbers of fermionic and bosonic fields.
Keeping in mind intrinsic non-localities of the supersymmetry transformations in the considered case, it is unlikely that the
supersymmetries in question can be directly related to the well-known supersymmetries of the super-extended WZW models.
For instance, ${\cal N}=(4,4)$ WZW models can be naturally described
in terms of the ``twisted-chiral'' ${\cal N}=(4,4)$ supermultiplets \cite{GHR,IK} with the off-shell field contents $(8 + 8)$,
in which 4 bosonic fields are auxiliary. At the same time, e.g., in the $n=3$ case, our full action \p{mod-act} in the gauge \p{Gau2}
contains $8$ fermionic and $6$ bosonic fields. To extend this set of fields to some off-shell multiplet,
we need to add at least two extra bosonic auxiliary fields, which does not match with the off-shell content of the twisted multiplet.
Also, the $n=5$ action in the same gauge  \p{Gau2} involves $16$ fermionic and $20$ bosonic fields, so we need at least
$4$ extra {\it fermionic} auxiliary fields to gain a genuine off-shell supersymmetry. It seems natural to analyze  these
problems within the appropriate off-shell superfield formalism, for instance, in the harmonic superspace approach \cite{HSS}
or its bi-harmonic generalization \cite{IS1} suitable just for ${\cal N}=(4,4), 2d$ systems. We hope to report on the results
of such a study elsewhere. Also, an interesting subject for the future consideration is the realization of the ${\cal N}=(8,8)$ supersymmetry,
found here in the case $n=5$ at the classical level, in the quantum PR $AdS_5\times S^5\,$ superstring theory. The coefficient
before $S_a$ in \p{mod-act} is properly changed in the quantum case \cite{RT}, which could give rise to the breaking
(or deformation) of the underlying supersymmetry.

The quantum theory based on the action $S_{tot}'$ can be plagued by massless $H$-valued ghosts\footnote{We are indebted to A. Tseytlin for this remark.}. Surprisingly, this ghost problem
can be evaded within our consideration by noting that the supersymmetry transformations \p{tranGau2} simultaneously provide
an invariance of the original action $S_{tot}$ which is free of such troubles (see Note added).

\section*{Acknowledgements}
We thank Maxim Grigoriev and Arkady Tseytlin for useful
correspondence and comments. E.I. is grateful to Marc Magro for early discussions and correspondence concerning the Pohlmeyer
reduction in superstring theories. Our work was supported by RFBR
grants 09-02-01209 and 09-01-93107.

\section*{Note added}
Since this paper has appeared in Archive, we became aware of two new papers \cite{new1,new2} treating similar
subjects. The authors of \cite{new2} have shown that some non-local  supersymmetry transformations (their eqs. (3.57), (3.58))
provide an invariance of the original (not modified) PR superstring action $S_{tot}\,$.
To make a contact with our consideration, we first notice that the transformations of \cite{new2} precisely coincide with ours
\p{tranGau2} under the following correspondence:
\bea
&&\Lambda =T\,, \;
\gamma\,\rightarrow\, g\,,\; \psi _+\rightarrow\frac{\Psi _R}{\sqrt{\mu}}\,,\; \psi _-\rightarrow\frac{\Psi _L}{\sqrt{\mu}}\,,
A_\pm\,\rightarrow  \, \frac{A_\pm}{\mu}\,,\; \sigma ^\pm\, \rightarrow\, \mu\sigma ^\pm\,,\; \partial _\pm\,\rightarrow\,
\frac{\partial _\pm}{\mu}\,, \;
\epsilon _+\rightarrow\sqrt{\mu}\epsilon _L\, \nn
&& q^\bot =\mu\partial _-^{-1}[\epsilon _L,\,(g^{-1}\Psi _Lg)^\bot]\,. \nonumber
\eea
Further, the choice of gauge \p{Gau2} in the action $S_{tot}$ can be equivalently interpreted as the following change of variables \cite{RT}:
\bea
g = u \tilde{g} \bar u^{-1}, \quad  \Psi_L = u \tilde{\Psi}_L u^{-1}\,, \quad
\Psi_R  = \bar u \tilde{\Psi}_R \bar u^{-1}\,, \nonumber
\eea
where $u$ and $\bar u$ are not assumed to be ${\bf 1}$. Then the transformations \p{tranGau2}, with all variables being replaced by those with tildas,
obviously leave invariant $S_{tot}$ and $S_a$ in $S_{tot}'$ separately, because the gauge degrees of freedom $u$ and $\bar u$,
and, hence, the additional term $S_a(u^{-1}\bar u)$ are not transformed at all. Then, coming back to the original variables in $S_{tot}$,
it is straightforward to find the non-local supersymmetry transformations which leave invariant  $S_{tot}$ also in these variables:
\bea
&& \delta_{\epsilon_L}\, g = g([T, [\Psi_R, \,\tilde\epsilon_L]] + \hat\delta \bar h)\,, \quad \delta_{\epsilon_L}\,\Psi_R
= [(g^{-1}D_+ g)^{||}, \,\tilde\epsilon_L] + [\Psi_R, \,\hat\delta \bar h]\,, \nn
&& \delta_{\epsilon_L}\,\Psi_L = \mu[T,\, g\tilde\epsilon_Lg^{-1}]\,, \quad \delta A_\pm = 0\,, \nonumber
\eea
where now $\hat\delta \bar h = \mu (D_-)^{-1}\,[\tilde\epsilon_L, (g^{-1}\Psi_Lg)^\bot]\,$.

\end{document}